\documentclass{nime-alternate} 

\usepackage{anonymize} 		   

\usepackage[utf8]{inputenc}

\begin{document}

\conferenceinfo{NIME'20,}{July 21-25, 2020, Royal Birmingham Conservatoire, ~~~~~~~~~~~~ Birmingham City University, Birmingham, United Kingdom.}

\title{Towards democratizing music production with AI---Design of Variational Autoencoder-based Rhythm Generator as a DAW plugin}

%
%
%
\label{key}
%

\numberofauthors{1} 

\author{
\alignauthor
\anonymize{Nao Tokui}\\
       \affaddr{\anonymize{Keio University}}\\
       \affaddr{\anonymize{5332 Endo, Fujisawa}}\\
       \affaddr{\anonymize{Kanagawa, Japan}}\\
       \email{\anonymize{tokui@sfc.keio.ac.jp}}
}


\maketitle

\begin{figure*}[h]
	\centering
		\includegraphics[width=1.0\textwidth]{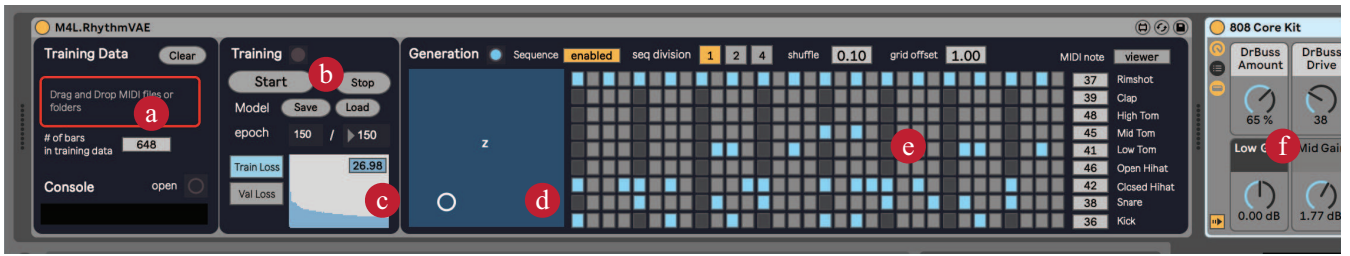}
	\caption{ M4L.RhythmVAE - screenshot of the proposed rhythm generator plugin. 
	\textcircled{\small a} Users can set training data by "drag and drop" MIDI files here. 
	\textcircled{\small b} Buttons for start/stop training. 
	\textcircled{\small c} Graphs for plotting training and validation losses.
	\textcircled{\small d} A standard XY pad for controlling input latent vectors for the trained VAE decoder. When the user moves a knob on the pad, the device instantly generates a new rhythm pattern, slightly different from the previous one, according to the 2-dimensional input vector represented on the XY pad.
	\textcircled{\small e} Visualization of the generated rhythm pattern. 	 
	\textcircled{\small f} The output of the device can be forwarded and played in standard drum software synthesizers of Ableton Live or external MIDI devices via MIDI output. 
	}
	\label{fig:device}
\end{figure*}

\begin{abstract}

There has been significant progress in the music generation technique utilizing deep learning. However, it is still hard for musicians and artists to use these techniques in their daily music-making practice. 

This paper proposes a Variational Autoencoder\cite{Kingma2014}(VAE)-based rhythm generation system, in which musicians can train a deep learning model only by selecting target MIDI files, then generate various rhythms with the model. The author has implemented the system as a plugin software for a DAW (Digital Audio Workstation), namely a Max for Live device for Ableton Live. 

Selected professional/semi-professional musicians and music producers have used the plugin, and they proved that the plugin is a useful tool for making music creatively.

The plugin, source code, and demo videos are available online\footnote{\anonymize{https://cclab.sfc.keio.ac.jp/projects/m4l-rhythmvae/}\label{foot:mainsite}}.

\end{abstract}




\section{Introduction}

Since the research of deep learning took off, researchers and artists have proved that deep learning models are competent for generating information contents, such as images and texts. Music is no exception. Many researchers have been working on applications of deep learning, especially architectures for time-series prediction such as Recurrent Neural Networks(RNNs), in music generation.

Historically, many of those researches focused on generating melodies, typically of piano, usually using recurrent neural networks(RNN) or their variant, Long-short term memory (LSTM)\cite{Hochreiter1997} (for example, \cite{Oore2018}). In the last couple of years, researchers had started working on models generating rhythms and bass lines and, in some cases, both of them alongside melodies, so that the model can generate music in the more comprehensive form with rhythm, bass, and melody parts\cite{Roberts2018b}. 

Few researches have tackled the music generation problem in the audio signal domain due to its significant complexity and computational requirements (One of a few exceptions is \cite{Manzelli2018}). Hence this paper focuses on music generation at the symbolic level. \cite{Briot} gives a more comprehensive view of current deep learning-based music generation techniques. 

These researches also tend to focus on the autonomous nature of the process: They design AI systems to autonomously generate a complete piece of music, or at least a part of music, according to a given input (a short sequence of notes as a seed, or a random input vector). Interventions of human musicians are usually not considered.  

Based on MusicVAE\cite{Roberts2018b} and GrooVAE\cite{Gillick2019}, Google Magenta team provided accompanying plugins for Ableton Live\footnote{\url{https://magenta.tensorflow.org/studio/ableton-live/}\label{foot:magenta}} so that musicians can use them to generate short melody and rhythm patterns. Although these plugins can be used in trial and error manner, the parameters provided by them are limited: "temperature" of the sampling process is the only parameter to affect the patterns to be generated. The user has to stop playing and hit a button to generate new patterns reflecting new parameter settings.


The generative models (in this case, decoders of Variational Autoencoders\cite{Kingma2014}) included in MusicVAE\cite{Roberts2018b} plugins were trained with millions of MIDI data provided as a publicly available Lakh MIDI Dataset (LMD)\cite{Raffel2016}. Users of the plugin do not have any control over MIDI files used in the training process, and subsequently, what kind of music patterns the model is capable of generating.  

By considering the limitations of current AI-based music generation tools as mentioned above, this paper proposes an easy-to-use deep learning-based music generation assistive tools, which musicians and music producers can use to generate novel music patterns and get new musical ideas from them.   

The tool proposed here is the following advantages:

\begin{enumerate}
  \item  Users of the tool can train their own AI model by "drag and drop" MIDI files they want to use in training onto the plugin. By doing so, they have loose control over the patterns they want the plugin to generate.  
  \item The tool provides direct access to the latent space of the trained model. Users can directory specify/modify input latent vectors and generate various rhythm patterns corresponding to the inputs.
  \item The plugin gives realtime feedback. When users move input vectors, the plugin instantly generates new rhythm patterns.  Users can listen to/try various rhythms by exploring the latent space.
\end{enumerate}

The author has implemented the tool as a Max for Live device(plugin) for Ableton Live\footnote{\url{https://www.ableton.com/live/max-for-live/}}---\textit{M4L.RhythmVAE}(Fig.\ref{fig:device}). Ableton Live\footnote{\url{https://www.ableton.com/live/}} is one of the most popular DAW(Digital Audio Workstation) softwares.  The plugin used Cycling'74 Max\footnote{\url{https://cycling74.com/}} and TensorFlow.js\footnote{\url{https://www.tensorflow.org/js/}}. TensorFlow.js enables us to integrate powerful machine learning capability into JavaScript runtime of Node for Max\footnote{\url{https://docs.cycling74.com/nodeformax/api/}} of Cycling'74 Max. 

On our website, the author provides a demo video\footnotemark{\ref{foot:mainsite}}. 

This paper focuses on generating rhythm patterns, keeping the production of electronic dance music in mind, but the same technique can be easily applied to melody generation and other music genres.

\section{Related Works}

\subsection{Rhythm generation}

Historically speaking, the majority of researches on music generation deal with melodies and harmonies. Out of 32 papers reviewed in \cite{Briot} (Chapter 7), only two papers handle rhythm as their main objective. However, researchers started experimenting with rhythm generation techniques recently. 

\cite{Choi} showed that a simple LSTM architecture trained with rhythm data encoded as strings of characters could successfully generate heavy-metal drum patterns. The paper used a binary representation of nine different drum components. For example, 100000000 and 010000000 represent kick and snare, respectively, and 110000000 means playing kick and snare simultaneously.  

In \cite{Makris}, Markris \textit{et al.} proposed an architecture with stacked LSTM layers conditioned with a Feedforward layer to generate rhythm patterns.
The Feedforward layer takes information on accompanying bass lines and the position within a bar. Training rhythm patterns in binary representation, similar to \cite{Choi}, are fed into LSTM layers. The system allows only five drum components (kick, snare, tom, hihat, cymbal).   

In \cite{Gillick2019}, Gillick \textit{et al.} employed Variational Autoencoder(VAE) model to generate new rhythm patterns. The paper also made use of the encoder-decoder architecture of VAE and proved that their proposed architecture could "humanize" quantized rhythms by stressing or weakening certain onsets and slightly shifting the timing of them. 

Their model handles a rhythm pattern represented as three matrices: onsets, velocities (strength of onsets), timing offsets from quantized timing (discussed more in detail in section \ref{sec:implementation}. Their VAE consists of layers of bidirectional LSTM used in \cite{Roberts2018b}, and the dimension of the latent space \textit{z} is 256. 

Gillick \textit{et al.} also provide Ableton Live plugin\footnotemark{\ref{foot:magenta}} based on the method, which allows users to generate new rhythms within the Ableton Live DAW environment. However, users cannot train their own models, and users' control over generated patterns is limited. The plugins do not operate in realtime either, i.e. users have to stop playing when they want to generate new patterns.

\subsection{Assistive tools}

There are not many literatures on the application of deep learning in the context of assistive tools for music composition since its researches are relatively new.

Hadjeres \textit{et al.} proposed DeepBach architecture for the generation of J.S. Bach chorals\cite{Hadjeres2016}. The architecture combines two LSTMs and two Feedforward layers. The user interface of the system was also implemented as a plugin for the MuseScore music editor. It allows the user to select generated chorales and control the progressions interactively. 

In \cite{Vogl2017}, Vogl \textit{et al.} proposed a concept of an "intelligent drum machine" utilizing a rhythm generation model based on restricted Boltzmann machines (RBMs) \cite{Hinton2006}. On their tablet interface, the user can input their own rhythm pattern (seed) in a grid UI and generate its variations by turning a knob. To generate variations of the seed pattern, first, the seed pattern is entered into the visible layer of the RBM. Variations for every instrument are generated at once using Gibbs sampling. The rhythm patterns are handled as sequences 0s and 1s: 0 means there is an active note, 1 means no active note.

The generated variations are sorted according to their similarity to the seed and number of active onsets. The authors used a variance of the Hamming distance assessed in \cite{TOUSSAINT2006})  as a similarity measure. Finally, patterns other than the 16 most similar ones in both groups are discarded. In this way, the system allows the user to explore the variations from sparse patterns to dense ones, from similar ones to different ones, by turning a knob UI. Users can also play generated rhythms via Ableton's Link\footnote{\url{https://www.ableton.com/link/}} technology as MIDI output. 

Although the concept and goal is similar to the ones presented in this paper,  the system in  \cite{Vogl2017} does not allow users to train their model, and its representation of rhythm is limited.

\section{Implementation} 
\label{sec:implementation}


\begin{figure}[t]
	\centering
	\includegraphics[width=0.85\columnwidth]{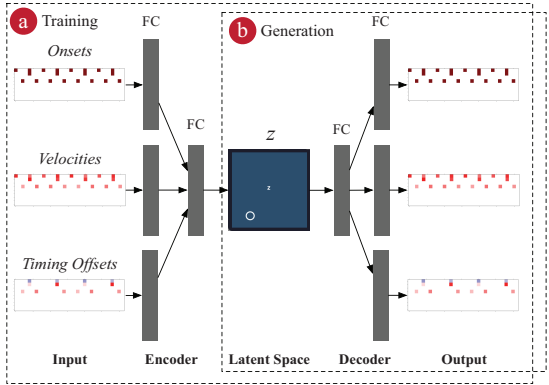}
	\caption{Overview of the VAE architecture used in the plugin. (FC: Fully-connected layer)}
	\label{fig:architecture}
\end{figure}

\begin{figure*}[htb]
	\centering
		\includegraphics[width=0.9\textwidth]{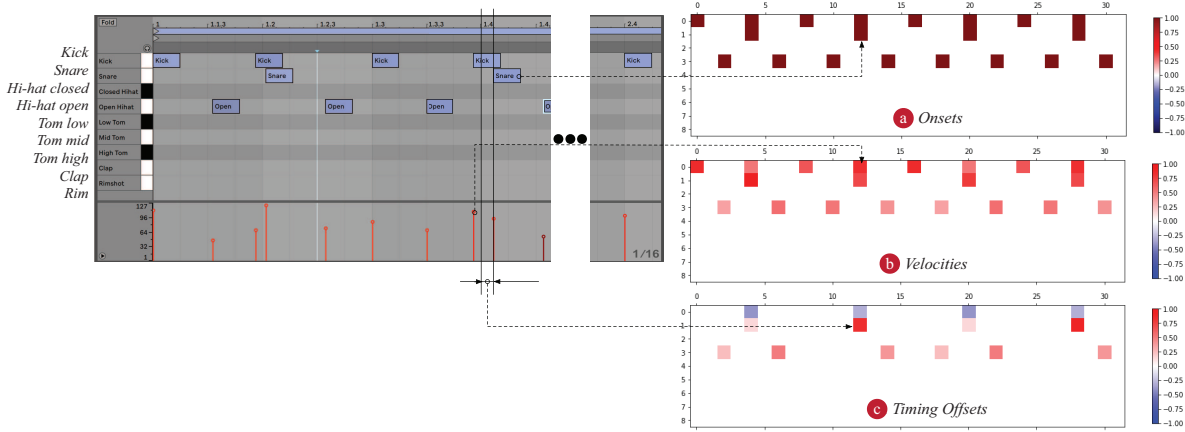}
	\caption{A sample rhythm pattern (left) and its encoded matrices (right).}
	\label{fig:encoding}
\end{figure*}

The generative model the plugin used here is a Variational Autoencoder(VAE)\cite{Kingma2014}. The author implemented the training process of the VAE model in the Max for Live device(plugin). The structure of the VAE is similar to \cite{Gillick2019}. The encoder encodes batches of three input matrices, which represent drum onsets, velocity (strength) of onsets, and timing offset from the given grid(Fig.\ref{fig:architecture}). Once the plugin finished training with user-provided MIDI files, it uses the decoder of the VAE to generate rhythm patterns(Fig.\ref{fig:architecture} \textcircled{\small b}).

The minimum time step of the drum onset is the 16th note; the encoder quantizes the timing of every onset to the timing of the nearest 16th note. Onsets are represented as {0, 1}(Fig.\ref{fig:encoding} \textcircled{\small a}). The encoder also normalizes the MIDI velocity values of onsets [0, 127] to [0., 1.](Fig.\ref{fig:encoding} \textcircled{\small b}). The timing offset value [-1.0, 1.0) represents the relative distance from the nearest 16th note. -1.0 indicates the given onset is a 32nd note ahead from the exact timing of the nearest 16th note(Fig.\ref{fig:encoding} \textcircled{\small c}). 

The author chose nine typical drum sounds (\textit{Kick, Snare, Hi-hat closed, Hi-hat open, Tom low, Tom mid, Tom high, Clap, Rim}). The author also specified mapping from MIDI note numbers in General MIDI (GM) convention\cite{MIDIAssociation1991} to the selected nine drums. The VAE model encodes and decodes two bar-long rhythm patterns. Since onset timings are quantized in 16th notes, the input and output of the model is 9 x 32 matrix.  

The main difference from the architecture proposed in \cite{Gillick2019} is that the plugin adopts simple fully-connected Feedforward layers instead of bidirectional LSTMs in favor of faster training on CPU environment of TensorFlow.js in Node for Max. More precisely, both encoder and decoder have two layers of Feedforward layers with 512 nodes with batch normalization\cite{Ioffe2015} and LeakyReLU activation. 

The VAE model has a 2-dimensional latent space; the encoder encodes input matrices into 2-dimensional vectors, and the decoder reconstructs the input matrices from these 2D vectors. The author deliberately reduced the dimension of the latent space down to 2, from 256 of the model proposed in \cite{Gillick2019}, so that users of the plugin can directly control the latent vector by standard XY pad UI (Fig.\ref{fig:device} \textcircled{\small d}). 


\begin{figure}[h]
	\centering
	\includegraphics[width=0.9\columnwidth]{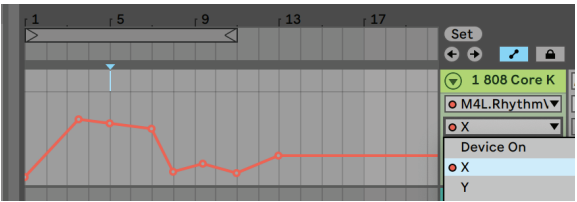}
	\caption{Automation of the progression of input latent vector for VAE decoder.}
	\label{fig:automation}
\end{figure}

When users use the plugin, they can select MIDI files they want to use, "drag and drop" these files(Figure.\ref{fig:device} \textcircled{\small a}). The plugin expects MIDI files conformed in General MIDI convention\cite{MIDIAssociation1991}, e.g., MIDI note number 36 corresponds to \textit{Bass Drum}, 40 is \textit{Electronic Snare}. It ignores channels other than channel \#10, which is dedicated to a drum track in GM Format. Every note onset in channel \#10 is mapped to one of 9 drums according to the pre-defined drum mapping.


Once the plugin loaded MIDI files, users of the plugin can start training a VAE model by just clicking a start button. Training losses and validation losses are plotted on respective graphs. Once the training for user-specified epochs finished, the user can use the decoder of the trained VAE to generate rhythms by moving a knob in XY-pad representing the 2D latent space of the VAE. 

The internal sequencer of the plugin plays the generated rhythm through MIDI output in sync with Ableton Live's global sequencer. The internal sequencer plays a drum onset slightly before or after the exact timing of 16th notes, depending on the "timing offset" output of the decoder.  

 When users gradually move the XY-pad knob, the decoder generates continuously changing rhythm patterns in realtime, so that the user can create musical progressions on the fly.  The movement of the knob of XY-pad---the transition in the latent space---can be recorded and played back afterward as a part of the automation mechanism of Ableton Live's sequencer(Fig.\ref{fig:automation}).

\section{Discussion}

\subsection{Real world test}

In November 2019, the author co-organized a workshop session with selected musicians from around the world as a part of MUTEK AI Music Lab Tokyo\footnote{\url{https://mutek.jp/en/news/archives/50}} and asked them to use the plugin. 

Feedbacks from the participants are mainly positive, and one participant stated that it gave him new musical ideas. Some of the participants used the plugin in realtime on stage during the final presentation\footnote{A short video documentation of these performances is available online: \url{https://youtu.be/RnlJF1YU6JU}}.

A more empirical analysis of the plugin and user questionaries are yet to be conducted. 

Some of the complete tracks made by participants of the workshop using rhythm patterns generated by the plugin are available online\footnotemark{\ref{foot:mainsite}}. 

\subsection{Model size and controllability}

One can argue that the VAE model the author adopted lacks the capacity of generalizing the wide range of different types of rhythm patterns since it has only 2-dimensional latent space in favor of the usability with the XY Pad. The smaller latent space leads to severe information bottleneck. Still, in the practical use case of this tool, where average users do not have millions of MIDI files, the author assumed the downside of the bottleneck is negligible. 

The author also observed that musicians do not need generic generative models. They usually prefer to have generative models, which enable them to explore specific kinds of music (in this case, rhythm patterns). (One of the participants even said, "I love overfitting.")

At the same time, if we further explore this idea of easy-to-use AI music tools for musicians and try generating more complex music patterns, then the limitation of the 2-dimensional latent space represented on an XY-pad will be problematic.  It poses a challenging question on how we can adopt latent spaces in higher dimensions and maintain the controllability for the user at the same time.  This problem shall be one of the next challenges the author will tackle.
 
\section{Conclusion}

To let musicians train and use their rhythm generation model, this paper proposed a VAE-based rhythm generator as a Max for Live device, a plugin for the popular DAW software, Ableton Live. Musicians have tested the plugin and found it useful as a tool to seek new ideas and extend their musical creativity. 

The author believes that this research and plugin are one of the first stepping stones towards more democratized use of AI in creative practices. In the near future, the author also hopes that every artist/creator should be able to train their own small AI models suited to their need, instead of using generic models someone else has trained, and explore new creative ideas with them.

\section{Acknowledgments}

\anonymize{This research was funded by the Keio Research Institute at SFC Startup Grant and Keio University Gakuji Shinko Shikin grant. The author wishes to thank Stefano Kalonaris and Max Frenzel for inspiration. The author also thanks Maurice Jones and Natalia Fuchs for organizing MUTEK AI Music Lab in Tokyo.} 
%
\bibliographystyle{abbrv}
\bibliography{nime-references} 


\end{document}